# A tunable hybrid qubit in a triple quantum dot

Bao-Chuan Wang,[1, 2] Gang Cao,[1, 2,*] Hai-Ou Li,[1, 2] Ming Xiao,[1, 2] Guang-Can Guo,[1, 2] Xuedong Hu,[3] Hong-Wen Jiang,[4] and Guo-Ping Guo[1, 2,*]

[1] *Key Laboratory of Quantum Information, University of Science and Technology of China,*

*Chinese Academy of Sciences, Hefei 230026, China*

[2] *Synergetic Innovation Center of Quantum Information & Quantum Physics, University of Science and Technology of China, Hefei, Anhui 230026, China*

[3] *Department of Physics, University at Buffalo, SUNY, Buffalo, New York 14260, USA*

[4] *Department of Physics and Astronomy, University of California at Los Angeles, California 90095, USA*

[*]Corresponding Authors: gcao@ustc.edu.cn; gpguo@ustc.edu.cn

## Abstract

We experimentally demonstrate quantum coherent dynamics of a triple-dot-based multi-electron hybrid qubit. Pulsed experiments show that this system can be conveniently initialized, controlled, and measured electrically, and has good coherence time as compared to gate time. Furthermore, the current multi-electron hybrid qubit has an operation frequency that is tunable in a wide range, from 2 to about 15 GHz. We provide qualitative understandings of the experimental observations by mapping it onto a three-electron system, and compare it with the double dot hybrid qubit and the all-exchange triple-dot qubit.

A fully controllable two-level system is an essential building block toward a scalable quantum computer. A gate-defined semiconductor quantum dot (QD), which can be manipulated electrically and fabricated using modern microelectronic technology, is considered an ideal platform for quantum computation [1–3]. Over the past decade, extensive progress has been achieved in the exploration of qubits based on the spin and charge degrees of freedom of QD-confined electrons [4–19]. An important objective in these studies is to improve the number of gate operations within the coherence time. Although spin qubits, which couple weakly to their environments, have long coherence times, their single-qubit operations are relatively slow [6,8,13,14]. In contrast, charge qubits can be manipulated quickly, albeit within a short coherence time, because of the strong electrical interaction [16–19].

A hybrid qubit of charge and spin is a practical scheme with which to achieve fast operations with a reasonably long coherence time [20, 21]. Relaxation and dephasing are suppressed efficiently through parallel energy levels that have different spin symmetries. Recently, experiments on the hybrid qubit in a Si/SiGe heterostructure have demonstrated fast coherent control using both pulse- and microwave-driven mechanisms [22–24]. By tuning the qubit parameters, Ramsey and Rabi decay times have been extended to more than 120 ns and 1 μs, respectively [25]. However, tunable operation frequency for this qubit design remains a difficult challenge because the qubit energy splitting is based on valley splitting in Si.

Adding quantum dots and/or electrons inevitably increase the size of the system Hilbert space, thus allowing a broader search for an optimal qubit encoding scheme that is both controllable and coherent [26]. For instance, a triple quantum dot with multiple interacting electrons has a highly tunable energy structure. All-exchange qubit based on three electrons in a triplet dot is one such example [27-30]. We have also studied a tunable hybrid qubit in a five-electron GaAs double quantum dot by taking advantage of the asymmetry-split excited states in one of the dots [31]. In short, increasing the number of electrons and/or quantum dots allows tuning of the mixture between charge and spin degrees of freedom, therefore provides a potential path to realizing an optimally encoded qubit.

In this Letter we report an experimental exploration to realize a controllable hybrid qubit in a linear triple quantum dot with asymmetric tunnel couplings in the multi-electron charge configuration of (6,2,3)-(7,1,3). We perform pulsed experiments to generate coherent oscillations that indicate the presence of quasi-parallel energy levels, the same favorable characteristics displayed by the double-dot hybrid qubit [20, 21]. More importantly, we find that the energy splitting of our hybrid qubit can be tuned conveniently in a wide range. By mapping our complex energy structure onto that of a three-electron triple dot, we provide a qualitative description of our experimental observations. The results could potentially lead to various applications, including a new encoding scheme that can be exploited on the triplet dot structure.

The linear triple dot we study is fabricated on a GaAs/AlGaAs heterostructure using a combination of photolithography and electron beam lithography, as is illustrated by the scanning electron microscopy image shown in Fig. 1(a). The two-dimensional electron gas is located about 90 nm below the surface of the heterostructure. The density and mobility of the two-dimensional electron gas are $2.0 \times 10^{11} cm^{-2}$ and $1.2 \times 10^5 cm^2 \cdot V^{-1} \cdot s^{-1}$, respectively. Metal gates (Ti-Au) D1–D7, H1 and H2 form the triple dot array, while gates U1–U3, together with H1 and H2, create a quantum point contact (QPC) as a sensor of the charge states of the quantum dot array. The transconductance response of the QPC channel is acquired using a lock-in amplifier with a small ac modulation voltage (typically 0.2 mV) applied to gate D1. A high-frequency signal, passing through a semi-grid coaxial cable from a room-temperature environment, is also applied to gate D1 using a bias tee. The device is cooled inside a refrigerator at a base temperature of 20 mK.

A typical charge stability diagram in Fig. 1(b) of our device shows three distinct charging line slopes (indicated by three dashed lines) corresponding to the three quantum dots [32], with the green (blue, black) dashed line indicating the charging line of the left (middle, right) dot. A careful investigation of the charge occupation indicates that the charge transition we study here (marked by the circle) is (6,2,3) to (7,1,3), where (l,m,r) denotes the electron numbers in the left, middle, and right dot, respectively. The smooth (sharp) anti-crossing marked by the rectangle (circle) implies that the middle and right (left) dots have strong (weak) tunnel coupling [33].

Our triple-dot system is always initialized in the ground state of charge configuration (6,2,3). We then use a single pulse, as shown in Fig. 1(c), to drive the system into the (7,1,3) configuration non-adiabatically. As a result, coherent oscillations are generated between the two lowest-energy states of the triple dot. Figure 1(c) shows the measured transconductance response of the QPC channel in the area of the charge transition between (7,1,3) and (6,2,3) for a short pulse duration of $t_p$ = 200 ps and a pulse height of $V_p$ = 400 mV, with a repetition rate of 40 MHz. The fringes between the two white dashed lines, marked by the yellow arrows, indicate Landau-Zener-Stückelberg interference in the (7,1,3) charge configuration, similar to what we observed before in other samples [15–17].

Figure 2(a) shows the transconductance of the QPC channel as a function of the pulse duration time $t_p$ and the detuning $\varepsilon_p$ of the starting point of the pulse [together with pulse height, it would determine how deep the system is pushed into the (7,1,3) region]. Two easily distinguishable patterns are indicated in pink and green in Fig. 2(b). The green-line pattern is right at the boundary separating the (7,1,3) and (6,2,3) region. It has the shape of a letter V on the side, and is characteristic of a lock-in measurement of a charge qubit, also known as a charge-qubit Larmor oscillation pattern [17, 23]. We thus conclude that the green pattern here results from charge oscillations between the left and middle dot, i.e. between (7,1,3) and (6,2,3) configurations for the three dots.

The pink-line oscillation pattern resides completely within the (7,1,3) region. It has nearly parallel fringes as we vary the detuning point within (7,1,3), indicating that the oscillation frequency depends only weakly on $\varepsilon_p$ (which is equivalent to the interdot detuning between the left and the middle dot). To extract the oscillation frequency's dependence on detuning, we perform a fast Fourier transform on the data of Fig. 2(a), and show the result in Fig. 2(c). Clearly, the oscillation frequency remains almost constant in a large range of the left-middle detuning. Since the oscillation frequency is given by the energy difference between the two relevant states accessible through the fast-pulse technique, we conclude that the energy splitting varies only slowly with the detuning, indicating a quasi-parallel energy spectrum.

Figure 2(d) is a line cut of the data along the red dashed line in Fig. 2(a), after subtracting a constant background. A fit in the form $Aexp[-(t-t_0)^2/T_2^{*2}]cos(\omega t+\theta)$ gives $T_2^* = 4.0\, ns$, shown by the red solid line in Fig. 2(d). While this dephasing time is much faster than a spin qubit, it is on par with an un-optimized double dot hybrid qubit [25].

We can make two observations from the experimental results obtained here. First, the coherent oscillations cannot be due to a hyperfine field gradient between the dots. The frequency of these oscillations is far too high to be generated by the small nuclear field. Second, the oscillations cannot be due to charge oscillations between the left and middle dots as in the case of a charge qubit, because the oscillation frequency of a charge qubit should be sensitive to $\varepsilon_p$, and the observed oscillations occur within the (7,1,3) region. We thus conclude with confidence that the coherent oscillations we have observed here are Larmor oscillations between the ground and first excited states within the (7,1,3) region.

Further investigation of the triple dot reveals another interesting feature, as shown in Fig. 3. Specifically, we find that the frequency of the coherent oscillation depends sensitively on the gate voltage on D6, which presumably influences the detuning of the right dot the most, while the coherent oscillation seems to occur in the left two dots where charge occupation has changed relative to the initial state. Figure 3(a) to 3(d) presents the measured QPC transconductance response for four different D6 gate voltages. The oscillation frequency changes more than three times, from about 2 to about 7 GHz, as the D6 voltage changes only 15%, from −0.5 to −0.42 V. In Fig. 4(a) we present a more complete dependence of the oscillation frequency on the D6 voltage. The frequency can be continuously increased to ~15 GHz by increasing the D6 voltage, at which point the oscillation is too fast to resolve.

The tunable oscillation frequency indicates that the energy difference between the parallel energy levels depends sensitively on the detuning of the right dot. On the surface this seems counter-intuitive, as the coherent oscillation is generated by changing the charge occupation in the left two dots, with the right dot occupation remaining constant. However, as we discuss in our theoretical model below, this unexpected dependence becomes quite reasonable when we realize that all three dots are coupled, and the relevant states are multi-electron states extended over all three dots.

We have extracted the coherence time of each frequency using data from the line cut at $\varepsilon = -80$ μeV. The results are presented using the blue dots in Fig. 4(a). Clearly, the two quantities display opposite trends: as the coherent oscillation frequency increases from 2 to 15 GHz, coherence time decreases from 6 to 1 ns.

Summarizing our experimental results, we observed coherent oscillations between the ground and first excited states within the (7,1,3) charge configuration with reasonably good coherent properties, and found that the oscillation frequency depends sensitively on the detuning of the right dot but insensitively on the detuning between the left and middle dot. The important questions we need to answer now are thus, first, what are the two states between which the coherent oscillations occur, and second, why the observed oscillation frequency has the particular detuning dependences.

A theoretical calculation of the eigenenergies and eigenstates of 11 electrons in a triple dot is not impossible. However, the strong Coulomb interaction and the resulting electron correlations mean that the eigenstates will not be single Slater determinants from single-particle states. Instead the electron eigenstates are always superpositions of multiple Slater deteminants, or "configurations". As such even an analytical solution would hardly give us any intuition on our problem. We thus focus on the qualitative physical picture and do not attempt to obtain a numerically accurate description of our system through a full-scale configuration interaction calculation.

There are two key features for the states of our triple dot, the charge distribution within each dot, and the spin symmetry of the many-electron states involved in our study. The former explains why we cannot repeat our experiments at lower charge occupation numbers. In other words, the electrons occupying the larger excited orbitals also see a lower tunnel barrier, making it much easier for us to observe correlated dynamics in the triple dot. The later feature, on spin symmetry, helps us map our multi-electron system onto a simpler system, and allows us to provide a qualitatively sensible physical picture for our experimental observations. Below we focus on this mapping.

Since Coulomb interaction is not spin dependent, each of the multi-electron eigenstates has a specific spin symmetry. For example, Hund's Rule dictates that in the ground state, 6 of the 7 electrons in the left dot should pair-up and form spin-singlets in the lowest three orbital states. Since our quantum dots are two-dimensional and nearly circular, these orbitals states should be the S and P Fock-Darwin states. Excitations from these close-shell states requires large energy, thus should contribute less to the low-energy states. Therefore the spin property of the 7 electrons in the right dot is determined by the lone outer-shell electron. The same argument can be made for the three electrons in the right dot. In all, we can thus argue that the spin symmetry of our triple-dot multi-electron states can be mapped to those for three electrons near the (1,1,1)-(0,2,1) charge transition.

To explain the basic features of our observations, we build a simple model based on the three outer-shell electrons [34]. The initial state of our experiment is the ground state in the (0,2,1) configuration with zero applied magnetic field, and the spin state (considering only the $S_z = +1/2$ component, without loss of generality since we do not consider spin-flip tunneling) is $|S\rangle_M|\uparrow\rangle_R$, where the subindex indicates the dot where the electrons are located [20-22]. The system is then driven to the (1,1,1) configuration with interdot exchange couplings $J_{LM}$ and $J_{MR}$ [26, 27]. We can use a spin Hamiltonian here because we have excluded charge dynamics between (1,1,1) and (0,2,1) so that we can focus on the spin dynamics within the (1,1,1) charge configuration.

The key to our observed coherent oscillation is that $J_{LM} \neq J_{MR}$ in our case, such that the total spin $S$ is not a good quantum number (while $S_z$ is). The ground and first excited states are thus mixtures of $|\uparrow\rangle_L|S\rangle_{MR}$ and $\sqrt{1/3}|\uparrow\rangle_L|T_0\rangle_{MR} -$

$\sqrt{2/3}\,|\downarrow\rangle_L |T_+\rangle_{MR}$ states [35], or the logical qubit states for the all-exchange qubit architecture [28-30]. In other words, when the system is suddenly driven from the initial (0,2,1) configuration into (1,1,1), both the ground and first excited states have finite probabilities of being occupied. The frequency of the ensuing coherent Larmor oscillation is at the energy splitting between these two states, given by $\sqrt{J_{LM}^2 + J_{LM}J_{MR} + J_{MR}^2}$.

The expression of the oscillation frequency hints at why we have the observed dependences on the two detunings. Specifically, at the limit when $J_{MR} \gg J_{LM}$, the oscillation frequency is approximately $J_{MR} + J_{LM}/2$, which is mostly determined by $J_{MR}$ and only slightly affected by $J_{LM}$. In other words, even though our pulse creates a charge transition between the left and middle dots, the frequency of the resulting coherent oscillation in the (1,1,1) regime is actually mostly determined by the stronger coupling between the middle and right dots, which is sensitive to the detuning between the middle and right dots controlled by D6 voltage. Conversely, the detuning between left and middle dots mostly affects $J_{LM}$, which only influences the oscillation frequency slightly. Thus our results seem insensitive to the left-middle detuning. In short, all the features of the results presented in Fig. 4(a) are expected within this model.

The increase in dephasing time in Fig. 4(b) can also be interpreted straightforwardly. Specifically, as $J_{MR}$ increases, dephasing of the coherent oscillation becomes increasingly susceptible to fluctuations in $J_{MR}$, so that decoherence effect of charge noise or other fluctuations becomes dominated by its influence on $J_{MR}$. Consequently, if we can identify a sweet spot for $J_{MR}$, it should be possible to have fast oscillations while enjoying good coherence properties [36-38].

In conclusion, we have demonstrated a tunable hybrid qubit in a triple quantum dot. The coherent oscillations we observe are results of free evolution between two energy levels insensitive to the left-middle detuning, while highly tunable by the right-middle detuning. Clearly, the addition of a third dot significantly increases the tunability of the qubit splitting compared to the original double-dot hybrid qubit. If a sweet spot can be found for the right-middle exchange coupling, this design has the potential of being widely tunable, highly coherent, and easily controllable. We hope that the results and discussions presented here stimulate further explorations into the quantum coherent dynamics in the few-electron regime for semiconductor quantum processors and nanoelectronics.

**Acknowledgements:** This work was supported by the National Key R & D Program (Grant No. 2016YFA0301700), the National Natural Science Foundation of China (Grants No. 11674300, No. 61674132, No. 11625419, No. 11575172, and No. 91421303), the "Strategic Priority Research Program (B)" of the Chinese Academy of Sciences (Grant No. XDB01030100), and the Fundamental Research Fund for the Central Universities. X.H. and H.W.J. acknowledge financial support by U.S. Army Research Office through Grants No. W911NF1210609 and No. W911NF1410346, respectively.

# References

[1] D. Loss and D. P. DiVincenzo, Phys. Rev. A 57, 120 (1998).

[2] G. Burkard, D. Loss, and D. P. DiVincenzo, Phys. Rev. B 59, 2070 (1999).

[3] I. Buluta, S. Ashhab and F. Nori, Rep. Prog. Phys. 74, 104401 (2011).

[4] T. Hayashi, T. Fujisawa, H. D. Cheong, Y. H. Jeong, and Y. Hirayama, Phys. Rev. Lett. 91, 226804 (2003).

[5] J. R. Petta, A. C. Johnson, C. M. Marcus, M. P. Hanson, and A. C. Gossard, Phys. Rev. Lett. 93, 186802 (2004).

[6] J. R. Petta, A. C. Johnson, J. M. Taylor, E. A. Laird, A. Yacoby, M. D. Lukin, C. M. Marcus, M. P. Hanson, and A. C. Gossard, Science 309, 2180 (2005).

[7] R. Hanson, L. P. Kouwenhoven, J. R. Petta, S. Tarucha, and L. M. K. Vandersypen, Rev. Mod. Phys. 79, 1217 (2007).

[8] K. C. Nowack, F. H. L. Koppens, Y. V. Nazarov, and L. M. K. Vandersypen, Science 318, 1430 (2007).

[9] M. D. Shulman, O. E. Dial, S. P. Harvey, H. Bluhm, V. Umansky, and A. Yacoby, Science 336, 202 (2012).

[10] X. Hao, R. Ruskov, M. Xiao, C. Tahan, and H. Jiang, Nat. Commun. 5, 3860 (2014).

[11] R. Brunner, Y. S. Shin, T. Obata, M. Pioro-Ladrière, T. Kubo, K. Yoshida, T. Taniyama, Y. Tokura, and S. Tarucha, Phys. Rev. Lett. 107, 146801 (2011).

[12] F. H. L. Koppens, C. Buizert, K. J. Tielrooij, I. T. Vink, K. C. Nowack, T. Meunier, L. P. Kouwenhoven, and L. M. K. Vandersypen, Nature 442, 766 (2006).

[13] B. M. Maune, M. G. Borselli, B. Huang, T. D. Ladd, P. W. Deelman, K. S. Holabird, A. A. Kiselev, I. Alvarado-Rodriguez, R. S. Ross, A. E. Schmitz, M. Sokolich, C. A. Watson, M. F. Gyure, and A. T. Hunter, Nature 481, 344 (2012).

[14] M. Veldhorst, J. C. C. Hwang, C. H. Yang, A. W. Leenstra, B. de Ronde, J. P. Dehollain, J. T. Muhonen, F. E. Hudson, K. M. Itoh, MorelloA, and A. S. Dzurak, Nat. Nano. 9, 981 (2014).

[15] K. D. Petersson, J. R. Petta, H. Lu, and A. C. Gossard, Phys. Rev. Lett. 105, 246804 (2010).

[16] G. Cao, H. O. Li, T. Tu, L. Wang, C. Zhou, M. Xiao, G. C. Guo, H. W. Jiang, and G.-P. Guo, Nat. Commun. 4, 1401 (2013).

[17] Z. Shi, C. B. Simmons, D. R. Ward, J. R. Prance, R. T. Mohr, T. S. Koh, J. K. Gamble, X. Wu, D. E. Savage, M. G. Lagally, M. Friesen, S. N. Coppersmith, and M. A. Eriksson, Phys. Rev. B 88, 075416 (2013).

[18] H.O. Li, G. Cao, G. D. Yu, M. Xiao, G. C. Guo, H. W. Jiang, and G. P. Guo, Nat. Commun. 6, 7681 (2015).

[19] D. Kim, D. R. Ward, C. B. Simmons, J. K. Gamble, R. B. Kohout, E. Nielsen, D. E. Savage, M. G. Lagally, M. Friesen, S. N. Coppersmith, and M. A. Eriksson, Nat. Nano. 10, 243 (2015).

[20] T. S. Koh, J. K. Gamble, M. Friesen, M. A. Eriksson, and S. N. Coppersmith, Phys. Rev. Lett. 109, 250503 (2012).

[21] Z. Shi, C. B. Simmons, J. R. Prance, J. K. Gamble, T. S. Koh, Y.-P. Shim, X. Hu, D. E. Savage, M. G. Lagally, M. A. Eriksson, M. Friesen, and S. N. Coppersmith, Phys. Rev. Lett. 108, 140503 (2012).

[22] D. Kim, Z. Shi, C. B. Simmons, D. R. Ward, J. R. Prance, T. S. Koh, J. K. Gamble, D. E. Savage, M. G. Lagally, M. Friesen, S. N. Coppersmith, and M. A. Eriksson, Nature 511, 70 (2014).

[23] Z. Shi, C. B. Simmons, D. R. Ward, J. R. Prance, X. Wu, T. S. Koh, J. K. Gamble, D. E. Savage, M. G. Lagally, M. Friesen, S. N. Coppersmith, and M. A. Eriksson, Nat. Commun. 5, 3020 (2014).

[24] D. Kim, D. R. Ward, C. B. Simmons, D. E. Savage, M. G. Lagally, M. Friesen, S. N. Coppersmith, and M. A. Eriksson, Npj Quantum Information 1, 15004 (2015).

[25] B. Thorgrimsson, D. Kim, Y. C. Yang, C. B. Simmons, D. R. Ward, R. H. Foote, D. E. Savage, M. G. Lagally, M. Friesen, S. N. Coppersmith, and M. A. Eriksson, arXiv:1611.04945 (2016).

[26] M. Russ and G. Burkard, arXiv:1611.09106 (2016).

[27] E. A. Laird, J. M. Taylor, D. P. DiVincenzo, C. M. Marcus, M. P. Hanson, and A. C. Gossard, Phys. Rev. B 82, 075403 (2010).

[28] J. Medford, J. Beil, J. M. Taylor, E. I. Rashba, H. Lu, A. C. Gossard, and C. M. Marcus, Phys. Rev. Lett. 111, 050501 (2013).

[29] J. Medford, J. Beil, J. M. Taylor, S. D. Bartlett, A. C. Doherty, E. I. Rashba, D. P. DiVincenzo, LuH, A. C. Gossard, and C. M. Marcus, Nat. Nano. 8, 654 (2013).

[30] N. Rohling and G. Burkard, Phys. Rev. B 93, 205434 (2016).

[31] G. Cao, H. O. Li, G. D. Yu, B. C. Wang, B. B. Chen, X. X. Song, M. Xiao, G. C. Guo, H. W. Jiang, X. Hu, and G. P. Guo, Phys. Rev. Lett. 116, 086801 (2016).

[32] L. Gaudreau, S. A. Studenikin, A. S. Sachrajda, P. Zawadzki, A. Kam, J. Lapointe, M. Korkusinski, and P. Hawrylak, Phys. Rev. Lett. 97, 036807 (2006).

[33] W. G. van der Wiel, S. De Franceschi, J. M. Elzerman, T. Fujisawa, S. Tarucha, and L. P. Kouwenhoven, Rev. Mod. Phys. 75, 1 (2002).

[34] See supplemental material for theoretical details.

[35] Y. P. Shim and C. Tahan, Phys. Rev. B 93, 121410 (2016).

[36] M. D. Reed, B. M. Maune, R. W. Andrews, M. G. Borselli, K. Eng, M. P. Jura, A. A. Kiselev, T. D. Ladd, S. T. Merkel, I. Milosavljevic, E. J. Pritchett, M. T. Rakher, R. S. Ross, A. E. Schmitz, A. Smith, J. A. Wright, M. F. Gyure, and A. T. Hunter, Phys. Rev. Lett. 116, 110402 (2016).

[37] F. Martins, F. K. Malinowski, P. D. Nissen, E. Barnes, S. Fallahi, G. C. Gardner, M. J. Manfra, C. M. Marcus, and F. Kuemmeth, Phys. Rev. Lett. 116, 116801 (2016).

[38] X. C. Yang, and X. Wang, arXiv: 1704.07975 (2017).

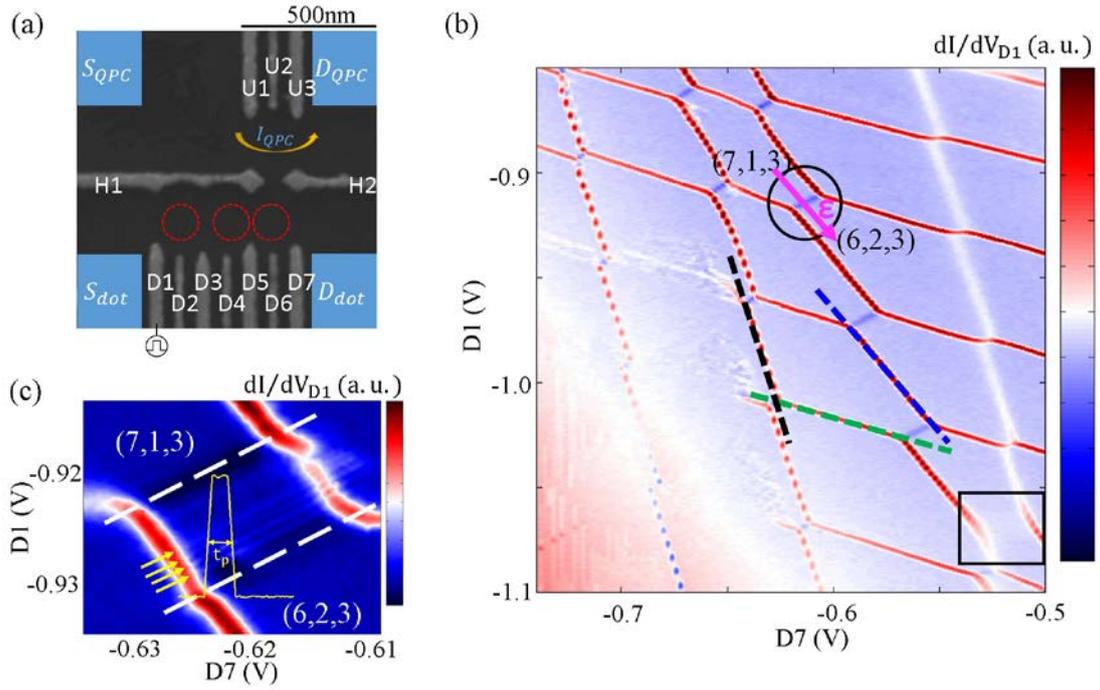

Fig. 1 (a) Scanning electron microscopy image of the device structure, where red dashed circles indicate the approximate quantum-dot positions. (b) Charge stability diagram of the triple quantum dot. Three dashed lines indicate three charging lines with different slopes. The solid circle indicates the area in which we perform our experiment. The pink arrow indicates the detuning direction. (c) Anti-crossing area indicated by the solid circle in (b), after applying a repeated pulse sequence. The inset schematically depicts the pulse sequence in the experiment.

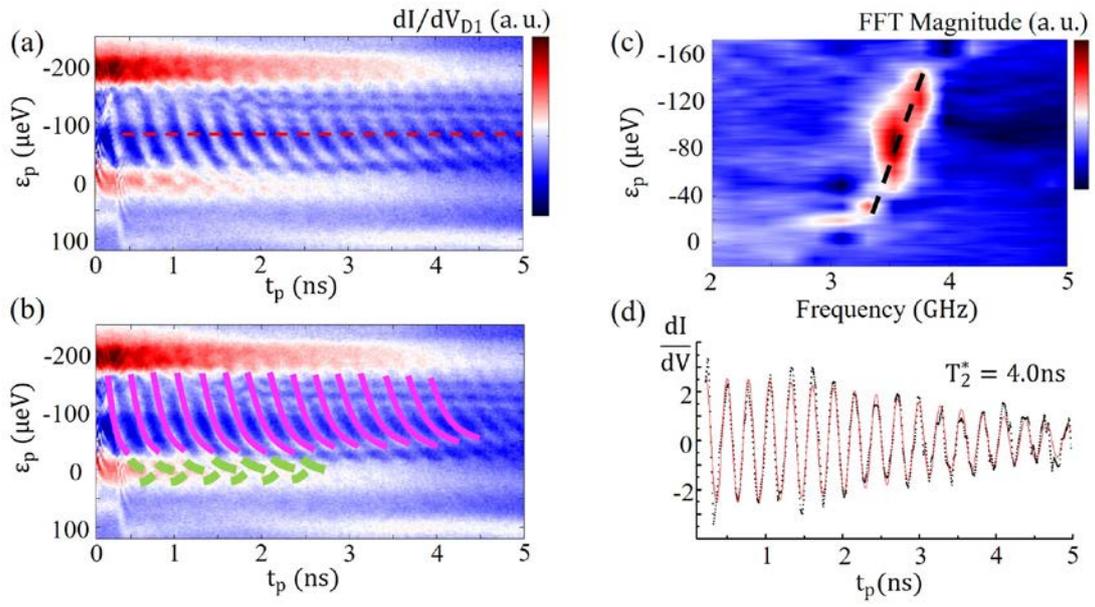

Fig. 2 (a) Coherent charge oscillations as a function of detuning $\varepsilon_p$ and pulse duration time $t_p$. (b) Two highlighted oscillation patterns for clarity of the same data in (a) using pink and green lines. (c) Fast Fourier transform of the data in (a). The dashed guideline indicates frequency variations. (d) Results for the dashed line in (a), after subtraction of a smooth background. The red solid line is a numerical fit, which yields $T_2^* = 4.0$ ns.

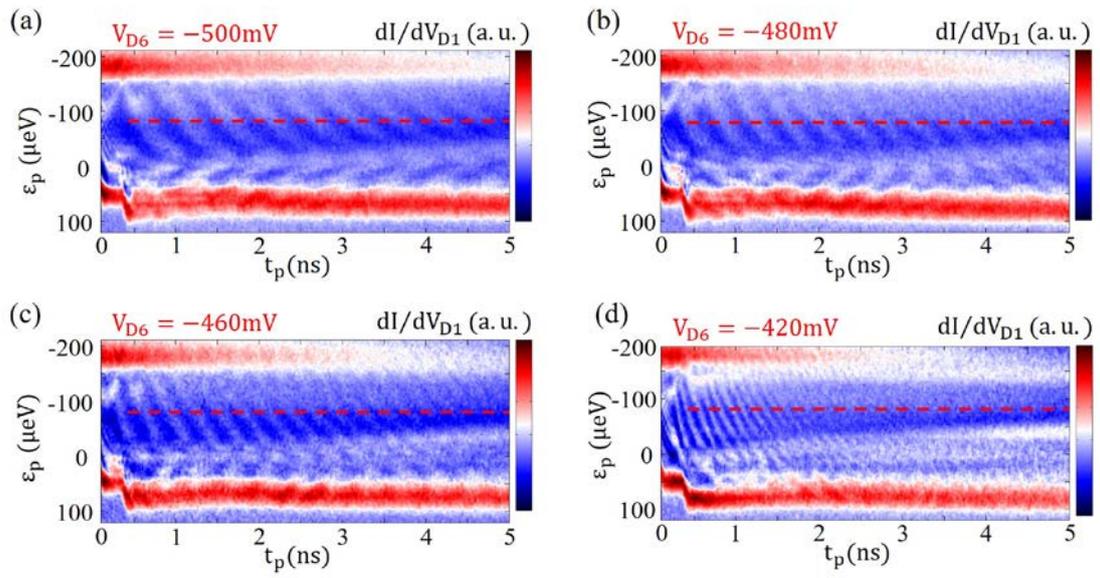

Fig. 3 Coherent charge oscillations as a function of detuning $\varepsilon_p$ and pulse duration time $t_p$ for four different values of $V_{D6}$. The oscillation frequency clearly increases with increasingly positive $V_{D6}$.

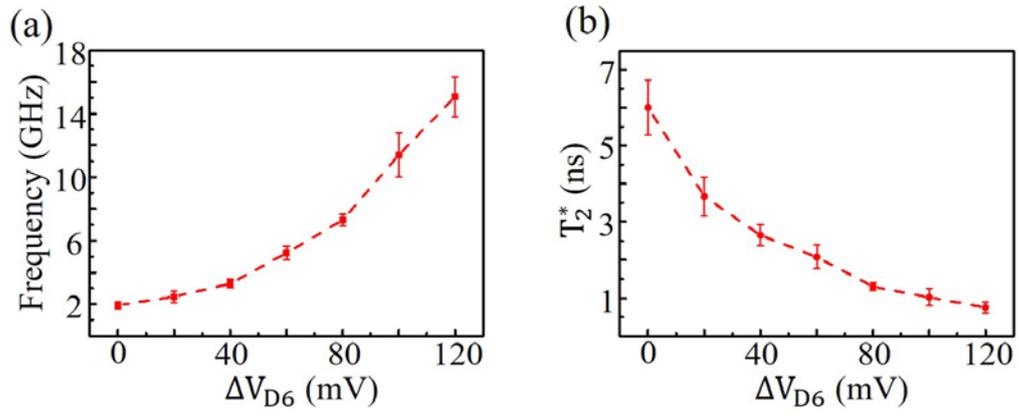

Fig. 4 (a) Charge oscillation frequency as a function of $V_{D6}$, extracted from the data of red dashed lines in Fig. 3 using the fast Fourier transform. (b) Extracted decoherence time using the same data in (a).

# Supplementary Material: A tunable hybrid qubit in a triple quantum dot

As we discussed in the main text, the qualitative features of many-electron states of the (7,1,3) and (6,2,3) charge configurations are similar to those of the (1,1,1) and (0,2,1) configurations. Since the observed coherent oscillations occur in the (7,1,3) regime, we attempt to understand these oscillations by examine the states of the equivalent (1,1,1) regime.

In the (1,1,1) charge configuration, the spin Hamiltonian for the three electrons is

$$H = J_{MR} \vec{S}_M \cdot \vec{S}_R + J_{LM} \vec{S}_L \cdot \vec{S}_M .$$

When $J_{MR} \gg J_{LM}$, we can treat the second part of the Hamiltonian as a perturbation, and start with the eigenstates of the first part of the Hamiltonian. Thus for basis of expansion we use product states of the single-spin eigenstates in the left dot and two-spin eigenstates of the middle and right dots ($S_{MR}$ and $T_{MR}$). In the $S_z = 1/2$ manifold, these states are:

$$|0\rangle = |\uparrow\rangle_L |S\rangle_{MR} = \frac{1}{\sqrt{2}} |\uparrow\uparrow\downarrow - \uparrow\downarrow\uparrow\rangle$$

$$|1\rangle = \frac{1}{\sqrt{3}} \left[ |\uparrow\rangle_L |T_0\rangle_{MR} - \sqrt{2} |\downarrow\rangle_L |T_+\rangle_{MR} \right] = \frac{1}{\sqrt{6}} |\uparrow\uparrow\downarrow - 2\downarrow\uparrow\uparrow + \uparrow\downarrow\uparrow\rangle$$

$$|Q\rangle = \frac{1}{\sqrt{3}} \left[ \sqrt{2} |\uparrow\rangle_L |T_0\rangle_{MR} + |\downarrow\rangle_L |T_+\rangle_{MR} \right] = \frac{1}{\sqrt{3}} |\uparrow\uparrow\downarrow + \downarrow\uparrow\uparrow + \uparrow\downarrow\uparrow\rangle .$$

Since $|0\rangle$ and $|1\rangle$ do not couple to $|Q\rangle$ even when we introduce $J_{LM} \vec{S}_L \cdot \vec{S}_M$ coupling, we can focus on $|0\rangle$ and $|1\rangle$.

The total Hamiltonian of (1,1,1) regime is:

$$\begin{aligned} H &= J_{MR} \vec{S}_M \cdot \vec{S}_R + J_{LM} \vec{S}_L \cdot \vec{S}_M \\ &= \frac{J_{MR}}{4} \vec{\sigma}_M \cdot \vec{\sigma}_R + \frac{J_{LM}}{4} \vec{\sigma}_M \cdot \vec{\sigma}_L \\ &= H_0 + \delta H . \end{aligned}$$

Here $H_0$ have eigenstates $|0\rangle$ and $|1\rangle$ with energies of $-\frac{3}{4}J$ and $\frac{J}{4}$ respectively. On the other hand, $\langle 0|\delta H|0\rangle = 0$, $\langle 1|\delta H|1\rangle = -\frac{j}{2}$ and $\langle 0|\delta H|1\rangle = \frac{\sqrt{3}}{4}j$. Thus we can write the Hamiltonian as:

$$H = H_0 + \delta H = \begin{pmatrix} -\frac{3}{4}J & \frac{\sqrt{3}}{4}j \\ \frac{\sqrt{3}}{4}j & \frac{3}{4}J - \frac{1}{2}j \end{pmatrix}.$$

The resulting eigenenergies of the $H$ are:

$$\lambda = \frac{1}{2}\left[-\frac{1}{2}(J+j) \pm \sqrt{J^2 - J\cdot j + j^2}\right] = \frac{1}{2}\left[-\frac{1}{2}(J+j) \pm J\sqrt{1-\frac{j}{J}+\left(\frac{j}{J}\right)^2}\right].$$

Since $J \gg j$, using Taylor expansion, the eigenenergies can be written as:

$$\lambda = \frac{1}{2}\left[-\frac{1}{2}(J+j) \pm \left(J - \frac{1}{2}j + \frac{3}{4}\frac{j^2}{J}\right)\right],$$

and the eigenstates of the total Hamiltonian in the (1,1,1) regime are: $|g\rangle = \alpha|0\rangle + \beta|1\rangle$ and $|e\rangle = -\beta|0\rangle + \alpha|1\rangle$ with $\beta = -\sqrt{3}\frac{j}{J}\alpha$.

When the voltage pulse is applied, it projects the initial state $|i\rangle = |S\rangle_M |\uparrow\rangle_R$ to $|g\rangle$ and $|e\rangle$. Firstly, we project the initial state onto $|0\rangle$ and $|1\rangle$:

$$\langle 0|i\rangle = \langle\uparrow|_L \langle S|_{MR} \cdot |S\rangle_M |\uparrow\rangle_R$$
$$= \frac{1}{\sqrt{2}}\langle\uparrow\uparrow\downarrow - \uparrow\downarrow\uparrow| \cdot \frac{1}{\sqrt{2}}|\uparrow\downarrow\uparrow - \downarrow\uparrow\uparrow\rangle = -\frac{1}{2}$$

$$\langle 1|i\rangle = \frac{1}{\sqrt{3}}\left[\langle\uparrow|_L \langle T_0|_{MR} - \sqrt{2}\langle\downarrow|_L \langle T_+|_{MR}\right] \cdot |S\rangle_M |\uparrow\rangle_R$$
$$= \frac{1}{2\sqrt{3}}\langle\uparrow\uparrow\downarrow - 2\downarrow\uparrow\uparrow + \uparrow\downarrow\uparrow|\uparrow\downarrow\uparrow - \downarrow\uparrow\uparrow\rangle = \frac{\sqrt{3}}{2}.$$

And in the basis of $|g\rangle$ and $|e\rangle$:

$$|i\rangle = (|g\rangle\langle g| + |e\rangle\langle e|)|i\rangle = \left(-\frac{\alpha}{2} + \frac{\sqrt{3}}{2}\beta\right)|g\rangle + \left(\frac{\beta}{2} + \frac{\sqrt{3}}{2}\alpha\right)|e\rangle.$$

Denote the energy splitting between the system excited state $|e\rangle$ and the ground state $|g\rangle$ by $\hbar\omega$, during the evolution in the (1,1,1) regime, the spin wave function can be written as:

$$|\psi(t)\rangle = \left(-\frac{\alpha}{2} + \frac{\sqrt{3}}{2}\beta\right)|g\rangle + e^{i\omega t}\left(\frac{\beta}{2} + \frac{\sqrt{3}}{2}\alpha\right)|e\rangle.$$

Here $\omega t$ indicate the phase accumulation between $|g\rangle$ and $|e\rangle$ during the pulse. At the end of the pulse, the system shifts back to the (0,2,1) regime, and $|\psi(t)\rangle$ is projected back to $|i\rangle$:

$$\langle i|\psi(t)\rangle = \left(-\frac{\alpha}{2} + \frac{\sqrt{3}}{2}\beta\right)^2 + e^{i\omega t}\left(\frac{\beta}{2} + \frac{\sqrt{3}}{2}\alpha\right)^2.$$

Since $\beta \ll \alpha$, we can make our estimate at the limit $\beta \sim 0, \alpha \sim 1$:

$$\langle i|\psi(t)\rangle = -\frac{1}{4} + \frac{3}{4}e^{i\omega t} \text{ and } \langle i|\psi(t)\rangle^2 = \frac{5}{8} + \frac{3}{8}\cos(\omega t).$$

This indicates that the return probability is between 1 and $\frac{1}{4}$.

We note that both $|g\rangle$ and $|e\rangle$ couple to the initial state $|i\rangle$, so that there is no spin blockade. Instead, the two states have different overlap with $|i\rangle$, thus have different rates of returning to $|i\rangle$. This difference in returning rates causes the system to stay in the (7,1,3) charge configuration with duration. In other words, the system spends different average time in (6,2,3) relative to (7,1,3), which causes a change in the average QPC signal.

In short, with $J \gg j$ we can explain the experimental observations and understand the fast measurement and initialization.